\documentclass[prd, showpacs,preprintnumbers,amsmath,amssymb,nofootinbib,aps]{revtex4}

\usepackage{bm}
\usepackage{amsfonts}
\usepackage{mathrsfs}
\usepackage{graphicx}
\usepackage{amsmath}
\usepackage{float}
\usepackage{braket}

\begin{document}

\newcommand{\hhat}[1]{\hat {\hat{#1}}}
\newcommand{\pslash}[1]{#1\llap{\sl/}}
\newcommand{\kslash}[1]{\rlap{\sl/}#1}
\newcommand{\lab}[1]{}
\newcommand{\iref}[2]{}
\newcommand{\sto}[1]{\begin{center} \textit{#1} \end{center}}
\newcommand{\rf}[1]{{\color{blue}[\textit{#1}]}}
\newcommand{\eml}[1]{#1}
\newcommand{\el}[1]{\label{#1}}
\newcommand{\er}[1]{Eq.\eqref{#1}}
\newcommand{\df}[1]{\textbf{#1}}
\newcommand{\mdf}[1]{\pmb{#1}}
\newcommand{\ft}[1]{\footnote{#1}}
\newcommand{\n}[1]{$#1$}
\newcommand{\fals}[1]{$^\times$ #1}
\newcommand{\new}{{\color{red}$^{NEW}$ }}
\newcommand{\ci}[1]{}
\newcommand{\de}[1]{{\color{green}\underline{#1}}}
\newcommand{\ke}{\rangle}
\newcommand{\br}{\langle}
\newcommand{\lb}{\left(}
\newcommand{\rb}{\right)}
\newcommand{\lbk}{\left[}
\newcommand{\rbk}{\right]}
\newcommand{\blb}{\Big(}
\newcommand{\brb}{\Big)}
\newcommand{\nn}{\nonumber \\}
\newcommand{\p}{\partial}
\newcommand{\pd}[1]{\frac {\partial} {\partial #1}}
\newcommand{\cd}{\nabla}
\newcommand{\cc}{$>$}
\newcommand{\bqa}{\begin{eqnarray}}
\newcommand{\eqa}{\end{eqnarray}}
\newcommand{\bqe}{\begin{equation}}
\newcommand{\eqe}{\end{equation}}
\newcommand{\bay}[1]{\left(\begin{array}{#1}}
\newcommand{\eay}{\end{array}\right)}
\newcommand{\eg}{\textit{e.g.} }
\newcommand{\ie}{\textit{i.e.}, }
\newcommand{\iv}[1]{{#1}^{-1}}
\newcommand{\st}[1]{|#1\ke}
\newcommand{\at}[1]{{\Big|}_{#1}}
\newcommand{\zt}[1]{\texttt{#1}}
\newcommand{\non}{\nonumber}
\newcommand{\m}{\mu}

\author{Morgan H. Lynch}
\email{mhlynch@uwm.edu}

\affiliation{Leonard E. Parker Center for Gravitation, Cosmology and Astrophysics, University of Wisconsin-Milwaukee,
P.O.Box 413, Milwaukee, Wisconsin USA 53201} 
\affiliation{Perimeter Institute for Theoretical Physics, 31 Caroline Street North, Waterloo, Ontario N2J 2Y5, Canada}

\title{Electron decay at IceCube}
\date{\today}

\begin{abstract}
In this paper we apply the formalism of Accelerated Quantum Dynamics (AQD) to the radiative stopping of highly relativistic electrons in ice. We compute the acceleration profile of the electron along with its lifetime to decay into a muon. The Planckian spectrum of the emitted muon along with the its generalized displacement law are presented and used to quantify the muons properties. The results predict the acceleration-induced decay of electrons at IceCube energies. The signal of electron decay at IceCube manifests itself as an excess of track topologies in an energy window accessible experimentally. This setting has the potential to probe the Unruh effect as well investigate the flavor content of cosmic ray neutrinos.
\end{abstract}


\maketitle

\section{Introduction}
The dynamics of quantum fields propagating and interacting in classical general relativistic backgrounds is described by quantum field theory in curved spacetime. Within its formalism is the natural extension of Minkowskian quantum field theory to curved spaces as well as the three canonical particle production mechanisms of the Parker [1], Hawking [2], and Unruh [3] effects. With the Unruh effect, there is a production of thermalized particles from an apparent horizon created by moving from an inertial into an accelerated reference frame. One can then use this thermal bath produced by the acceleration to induce particle transitions [4-9]. This transition rate of particles under acceleration is equivalent to the interaction rate of particles in a thermal bath at the accelerated temperature $t_{a} = \frac{a}{2\pi}$. AQD [9] is a formalism capable of analyzing a wide variety of these acceleration-induced processes and the observables which characterize them. Here we apply the formalism to the enormous deceleration felt by a charged particle, in the radiative regime, stopping in ice. Our analysis will focus specifically on the accelerated transition of electrons back into muons. We will show there exists an energy scale above which the electrons lifetime is smaller than the time it takes to exit the radiative regime and thus has a significant probability for the decay to occur. In the event of an electron decay, the energy of the emitted muon, as measured in the proper frame of the electron, has a generalized Planck distribution and displacement law that predicts the peak energy to be directly proportional to the accelerated temperature. When the muon is boosted back to the lab frame, there exists a large parameter space of energy which can be used to investigate the role of acceleration in the muon emission. In particular, there are kinematically forbidden regions where the energy of acceleration $E_{a}= \frac{xa\hbar}{2\pi c}$ can be investigated. Moreover, there exists the possibility of probing the Unruh effect [3] along with its similarities and differences with the radiative processes of the standard model [10]. The analysis is applied specifically to the energies probed by the IceCube detector. The signals of electron decay are translated into the signal topologies which characterize the data at IceCube. Evidence for electron decay will manifest itself via an increase in the number of track topologies in the energy window where the decay is expected to occur. Measurement of the shower to track ratio both in and out of the electron decay window also provides an alternative method to directly measure the incoming neutrino flavor content.

In this paper, Sec. II outlines the use of radiative energy loss in determining the deceleration profile of the particle. The relevant kinematic quantities which characterize the motion are also presented. Section III examines electrons and muons radiatively stopping in various materials and the energy scales which characterize their motion, i.e. radiation length, critical energy, accelerated temperature, etc. A particular emphasis is placed on an ice medium. In Sec. IV we compute the electron lifetime and compare it to the other time scales which characterize the motion. We show there exists an energy regime where the acceleration can be treated as time-independent and the electron has a statistically likely probability of decaying. Section V computes the muon spectrum and determines the most probable energy via application of the displacement law. We then boost the muon back into the lab frame and present the parameter space of the correlated electron and muon energies. Discussions on the Unruh effect and conservation of energy are included. Section VI outlines the signal topologies at IceCube and how they can be used to measure the presence of electron decay in the form of an excess of tracks in the detector. In Sec. VII we summarize the main conclusions of the paper. The natural units of $\hbar = c = k_{B} = 1$ are implemented throughout.

\section{Radiative deceleration Profile}

In this section we analyze the acceleration profile of a charged particle stopping in a dielectric medium. For ultra relativistic electrons, in the radiative regime, the energy loss is dominated by the emission of bremsstrahlung [11]. The ratio of radiative energy loss to the energy loss of collisions grows linearly with the electrons energy $E$. For energies above the critical energy, where the two energy loss mechanisms are equal, the radiative energy loss will dominate. For electrons in ice, the critical energy is $E_{c} \sim 79$ MeV. In this manuscript we are considering electron energies on the order of $E \sim 100$ TeV. Thus for our present concern the bremsstrahlung emission dominates over collision events by 6 orders of magnitude. This ensures a smooth energy loss and subsequent deceleration. The background medium may also induce a transition in an accelerated system, in this case the electron, by coupling to the lattice vibrations of the bulk crystalline matter. However, it has been shown [12] that an accelerated detector moving relativistically in a background thermal bath will undergo transitions dominated by the accelerated temperature. This implies that any coupling between the accelerated electron and vibrations within the ice, i.e. phonons, will be suppressed, and thus negligible, in the relativistic regime. The simultaneous emission of Larmor radiation, in addition to Unruh radiation, has also been investigated in the cases of electrons in vacuum as well as a plasma [13,14]. The angular distributions of the emitted radiation have certain regimes, namely forward and backward proper frame emission directions, where they do not overlap and interfere. This ensures the proposed process of electron decay will not be suppressed by interference with the emitted bremsstrahlung. Finally, the use of the interactions with bulk matter as a mechanism of deceleration to induce particle emission has been studied using laser pulses fired into a dielectric medium [15]. In this system the index of refraction, which is characterized by the laser frequency, material properties, and their interaction, is responsible for the deceleration. What is of particular importance is that the emitted photons were measured experimentally and thus provided confirmation of the theory. In regards to the present manuscript, this implies that the use of the interaction of the electron with the background ice, as a method of deceleration, will not spoil the acceleration-induced emission process. The bremsstrahlung itself already has a description from the point of view of the Unruh effect [16] and its presence alone may indeed a priori imply the validity the analysis. Thus, to develop the acceleration profile, we recall the standard energy loss [10] of charged particles in the radiative regime is given by

\bqa
\frac{dE}{dx} &=& - \frac{E}{x_{0}} \non \\
\Rightarrow E(x) &=& E_{0}e^{-x/x_{0}}.
\eqa 

The radiation length $x_{0}$ determines the distance traveled until the particle loses $e^{-1}$ of its energy. Here we will only consider ultra-relativistic particles with total energy is given by $E = m \gamma$. In this regime, the proper velocity $u$ can be approximated via $u \sim \gamma$. The above expression then gives the proper velocity as a function lab frame distance. Thus,

\bqa
u(x) = u_{0}e^{-x/x_{0}}.
\eqa

From this we can separate variables and determine the lab frame distance traversed as a function of proper time. Hence

\bqa
\frac{dx}{d\tau} &=& u_{0}e^{-x/x_{0}} \non \\
\int e^{x/x_{0}} dx &=& \int u_{0} d\tau \non \\
\Rightarrow x(\tau) &=& x_{0}\ln{(\tau / \tau_{0} + 1)}.
\eqa

We have defined $\tau_{0} = x_{0}/u_{0}$ and fixed the penetration into the medium at $x(\tau = 0)=0$. The proper velocity and acceleration as a function of proper time can then be easily obtained by differentiation. Thus 

\bqa
u(\tau) &=& \frac{u_{0}}{(\frac{\tau}{\tau_{0}} + 1)} \non \\
a(\tau) &=& \frac{a_{0}}{(\frac{\tau}{\tau_{0}} + 1)^{2}}.
\eqa 

We have dropped the sign associated with the deceleration since we are only concerned with the magnitude. Also, we assume the electron propagates semi-classically along the above derived trajectories, thus giving a well defined world line. The acceleration scale is set by the quantity $a_{0} = u_{0}^{2}/x_{0}$ which is both a property of the material as well as the initial energy. Recalling for ultra-relativistic velocities $u_{0} \sim \gamma = E_{0}/m$, we list the relevant kinematic quantities which set the scale of our system as follows:

\bqa
a_{0} &=& \frac{E_{0}^{2}}{x_{0}m^{2}} \non \\
u_{0} &=& \frac{E_{0}}{m} \non \\
\tau_{0} &=& \frac{x_{0}m}{E_{0}}.
\eqa

The critical energy $E_{c}$ of a material is the scale below which the radiative processes no longer dominate. As such, we can also define the low acceleration cutoff $a_{c} = \frac{E_{c}^{2}}{x_{0}m^{2}}$ below which we will no longer be able to use the radiative stopping power as given in Eqn. (1). We can also define a critical time $\tau_{c}$ which characterizes the amount of proper time it takes for a particles energy to reduce to the critical energy and exit the radiative regime. Inverting the proper acceleration in Eqn. (4) and evaluating the proper time at the critical acceleration we have

\bqa
\tau_{c}  &=& \tau_{0}\lbk \lb \frac{a_{0}}{a_{c}}\rb^{1/2} -1\rbk \non \\
&=& \frac{mx_{0}}{E_{0}}\lbk  \frac{E_{0}}{E_{c}} -1\rbk \non \\
&\sim & \frac{mx_{0}}{E_{c}}.
\eqa

In the last line we assumed $E_{0} \gg E_{c}$. Finally, the radiative energy loss of electrons will dominate up until the energy scale where the LPM effect begins to spoil the deceleration [17]. This suppression of photons with wavelengths of the order of the distance between scattering sites begins to diminish the energy loss near the LPM energy scale $E_{LPM}$. The above deceleration profile will therefore only be valid for incident electrons with energy $E_{0} < E_{LPM}$. Under the LPM threshold, the relevant parameters that characterize the system can be written in terms of the incident particle energy, incident particle mass, the radiation length, and critical energy of the decelerating material. We will now analyze the relevant kinematic quantities specifically for electrons and muons in various materials with a particular emphasis on ice.

\section{Energy scales of electrons in Ice}

In the previous section we outlined the relevant parameters which characterize our decelerated motion. Table I contains the relevant kinematic parameters for various materials. Moreover, the focus of this paper will be the excitation of electrons back into muons and, as we shall see in the next section, the energy scale of the acceleration is best written in terms of the muon mass. Defining the dimensionless acceleration $\tilde{a}_{0} = a_{0}/m_{\m}$, and likewise for the critical acceleration, we can look at the relevant energy scales for various materials. To compute the accelerations, we assume an incident electron energy of $E_{0} = 100$ TeV.

\begin{table}[H]
\centering
\begin{tabular}{ cccccc}
\hline \hline \\ [-2.0 ex]
Material~&~$x_{0}$~[cm]~&~$E_{c}$~[MeV]~&~$E_{LPM}$~[TeV]~&~$\tilde{a}_{c}$~[$\times 10^{-10}$]~&~$\tilde{a}_{0}$ \\
\hline \\ [-2.0 ex]
Uranium & 0.317 & 6.65 & 2.39 & 1 & 22700 \\
\hline \\ [-2.0 ex]
Lead & 0.561 & 7.43 & 4.32 & .708 & 12830 \\
\hline \\ [-2.0 ex]
Iron & 1.757 & 21.7 & 13.5 & 1.93 & 4096  \\
\hline \\ [-2.0 ex]
Ice & 39.31 &  78.6 & 303  & 1.13 & 183.1 \\
\hline \\ [-2.0 ex]
\hline
\end{tabular}
\caption{The radiation length, critical energy, LPM energy, critical acceleration, and initial acceleration for 100 TeV electrons in various materials.}
\end{table}

Focusing specifically on ice, Table II contains the dimensionless acceleration scale for different incident electron energies. For completeness, we include the accelerated temperature $t_{a}$ which also characterizes the overall energy scale of the decay process. The dimensionless acceleration enables these acceleration-induced processes to occur and regimes where $\tilde{a}_{0}\gg 1$ are where these effects are strongest. Moreover, in the interest of measuring these effects, ice is the easiest since it is transparent to light and thus facilitates measuring the energy of the system via bremsstrahlung. It is worth noting that for incident electron energies $E_{0} \lesssim 500$ TeV the accelerated temperature $t_{a}$ is at, or below, the weak scale $t_{a} < m_{W^{\pm}}$. We leave an analysis that includes the weak gauge bosons at higher energy scales for future work.

\begin{table}[H]
\centering
\begin{tabular}{ ccc}
\hline \hline \\ [-2.0 ex]
$E_{0}$~[TeV]~&~$\tilde{a}_{0}$~&~$t_{a}$~[GeV] \\
\hline \\ [-2.0 ex]
10 & 1.831 & .031\\
\hline \\ [-2.0 ex]
50 & 45.77 & .77 \\
\hline \\ [-2.0 ex]
100 & 183.1  & 3.08 \\
\hline \\ [-2.0 ex]
300 & 1637 & 27.5 \\
\hline \\ [-2.0 ex]
500 & 4577 & 77\\
\hline \\ [-2.0 ex]
\hline
\end{tabular}
\caption{The acceleration scale and temperature for electrons at various energies in ice.}
\end{table}

It is also worth tabulating the same parameters for muons in various materials. Table III looks at the kinematic parameters of 100 TeV muons moving through matter in the radiative regime. Here we use the same dimensionless acceleration written in terms of the muon mass. It should be noted that muons have a much larger radiation length and critical energy than electrons. Moreover the LPM energy scale for muons is a factor $\sim 10^{10}$ larger than electrons and is therefore negligible [10]. This allows us to analyze certain energy regimes where the electron is in the radiative regime and decay is most probable but the final state muon does not decay. This will better enable us to accurately analyze these processes.

\begin{table}[H]
\centering
\begin{tabular}{ ccccc}
\hline \hline \\ [-2.0 ex]
Material~&~$x_{0}$~[m]~&~$E_{c}$~[GeV]~&~$\tilde{a}_{c}$~[$\times 10^{-10}$]~&~$\tilde{a}_{0}$~[$\times 10^{-5}$] \\
\hline \\ [-2.0 ex]
Uranium & 26.69 & 128 & 1.046 & 6.387 \\
\hline \\ [-2.0 ex]
Lead & 48.03 & 141 & .706 & 3.5497 \\
\hline \\ [-2.0 ex]
Iron & 152.5 & 347 & 1.346 & 1.1178  \\
\hline \\ [-2.0 ex]
Ice & 3120 & 1031 & .58074 & .054634 \\
\hline \\ [-2.0 ex]
\hline
\end{tabular}
\caption{The radiation length, critical energy, critical acceleration, and initial acceleration for 100 TeV muons in various materials.}
\end{table}

Then, for ice, Table IV contains the dimensionless acceleration and accelerated temperature for muons. It should be noted that the acceleration scale and temperature are substantially smaller for the muon than for the electron. This is the result of having a much larger radiation length and critical energy. Moreover this coincidence helps to ensure that the accelerated decay processes we are looking for are more prominent for electrons than for muons. It should also be noted that the accelerated temperature for muons is substantially lower than the weak scale and the acceleration scale $\tilde{a}_{0} \ll 1$.  

\begin{table}[H]
\centering
\begin{tabular}{ ccc}
\hline \hline \\ [-2.0 ex]
$E_{0}$~[TeV]~&~$\tilde{a}_{0}$~[$\times 10^{-7}$]~&~$t_{a}$~[KeV] \\
\hline \\ [-2.0 ex]
100 & 5.4634 & .009\\
\hline \\ [-2.0 ex]
200 & 21.8536 & .038\\
\hline \\ [-2.0 ex]
400 & 87.4146 & .147 \\
\hline \\ [-2.0 ex]
800 & 349.658 & .588\\
\hline \\ [-2.0 ex]
1000 & 546.341 & .919\\
\hline \\ [-2.0 ex]
\hline
\end{tabular}
\caption{The acceleration scale and temperature for muons in ice at various energies.}
\end{table}

In this section we tabulated the various parameters that characterize the deceleration of both electrons and muons in various materials. A particular emphasis was placed on ice so we can search for these processes at the energy scales of IceCube [18]. The transparency of ice to light, as well as the distinct topologies of electrons and muons will help characterize the energies as well as the type of event that occurs [19]. We saw that  there isn't sufficient stopping power in the ice to bring the muons to the necessary acceleration scale. The electrons however do have sufficient deceleration. It is this asymmetry that will help define a possible event. 

\section{The Electron lifetime}

The formalism of AQD [9] is capable of computing a wide variety of observables of acceleration-induced interactions. In this section we compute the lifetime of an accelerated electron. This process is essentially the inverse beta decay of muons rather than protons. The relevant transition is given by
\bqa
e^{\pm} \rightarrow_{a} \mu^{\pm} + \bar{\nu}_{\mu} + \nu_{e}.
\eqa 

Here, we will model all final state particles as being massless since the accelerated temperature of the system is much larger than the muon and neutrino rest masses. The neutrinos can be easily approximated as massless particles. The acceleration-induced transition rate for arbitrary $n$-particle final state multiplicity [8, 9] is given by 

\bqa
\Gamma(\Delta E,a,n) = G_{n}^{2}\lb\frac{\Delta E}{\pi} \rb^{2n-1} \frac{1}{(4n -2)!!} \prod_{k = 0}^{n-1}\lbk 1  + k^{2} \lb \frac{ a}{\Delta E} \rb^{2} \rbk  \frac{1}{e^{2\pi\Delta E/|a|} - 1}.
\eqa

We note this formalism does include time dependent accelerations. Now, to model the electron decay we note there will be three final state particles, i.e. $n = 3$ for the muon and two neutrinos. Moreover, since the electron and neutrino masses are much smaller than the muon, the transition energy will be the muon mass, i.e. $\Delta E =  m_{\m} + m_{\nu_{\m}} + m_{\nu_{e}} - m_{e} \approx m_{\m}$. The electron decay rate is then given by

\bqa
\Gamma_{e}(\tilde{a}) &=& G^{2}\frac{m_{\m}^{5}}{3840 \pi^{5}} \frac{1+ 5\tilde{a}^{2}+ 4\tilde{a}^{4}}{e^{2\pi /|\tilde{a}|}-1}.
\eqa 

Note, we have written the decay rate in terms of the dimensionless acceleration $\tilde{a} = a/m_{\m}$. For transitions between the electron and muon the energy scale is set by the muon mass. We can fix the coupling $G^{2}$ via taking the inertial limit of the acceleration-induced muon decay and use the detailed balance to import the coupling for the electron excitation [8]. The acceleration-dependent muon decay rate can be obtained from Eqn. (9) simply by letting $m_{\m} \rightarrow -m_{\m}$, i.e. we transition down in energy. The daughter electron and neutrinos can be considered massless which again yields $n = 3$. This technique is equivalent to using the detailed balance of the two process at thermal equilibrium. Also note that this requires $|\tilde{a}| \rightarrow -|\tilde{a}|$. The acceleration-dependent muon decay rate [4,8] is then given by

\bqa
\Gamma_{\m}(\tilde{a}) &=& G^{2}\frac{m_{\m}^{5}}{3840 \pi^{5}} \frac{1+ 5\tilde{a}^{2}+ 4\tilde{a}^{4}}{1- e^{-2\pi /|\tilde{a}|}}.
\eqa 

The inertial muon decay rate [10] is well known to be

\bqa
\lambda_{\m} = \frac{G^{2}_{f}m^{5}_{\mu}}{192 \pi^{3}}.
\eqa

We can then fix the coupling by taking the inertial limit of the acceleration-dependent muon decay rate and matching it to the inertial rate. Thus

\bqa
\lim_{a \rightarrow 0} \Gamma_{\m}(\tilde{a}) &=& \lambda_{\m} \non \\
\Rightarrow G^{2} &=& \frac{35}{8} \pi^{2} G^{2}_{f}.
\eqa

Now that we have fixed the coupling, we have the acceleration-induced electron decay rate is given by

\bqa
\Gamma_{e}(\tilde{a}) &=& \frac{G^{2}_{f}m^{5}_{\m}}{192 \pi^{3}}  \frac{1+ 5\tilde{a}^{2}+ 4\tilde{a}^{4}}{e^{2\pi /|\tilde{a}|}-1}.
\eqa

If we recall the canonical muon lifetime is given by $\tau_{\m} = \frac{192 \pi^{3}}{G^{2}_{f}m^{5}_{\m}} = 2.184 \;\m s$, we can determine the lifetime by merely reciprocating the decay rate. Hence

\bqa
\tau_{e}(\tilde{a}) &=& \tau_{\m} \frac{e^{2\pi / \tilde{a}}-1 }{1+ 5  \tilde{a}^{2} + 4 \tilde{a}^{4}}.
\eqa

To better understand the decay process we must examine the various time scales of the system. We know the critical time $\tau_{c}$ from Eqn. (6) determines how long the electron would take to exit the radiative regime under the assumption of an initial energy much larger than the critical energy. For electrons in ice the critical time is given by $\tau_{c} = \frac{m_{e}x_{0}}{E_{c}} = 8.52\times 10^{-12}$ s. Furthermore for lifetimes smaller than the time scale $\tau_{0}$ from Eqn. (5), we can assume the acceleration to be constant. For our system the time scale is given by $\tau_{0} = \frac{m_{e}x_{0}}{E_{0}} = \frac{6.7\times 10^{-16}}{E_{0}}$ s. Here we are measuring $E_{0}$ in units of TeV. Finally, we must examine the lifetime $\tau_{e}$ in the limit of high acceleration, i.e. $a \gg m_{\m}$. This will enable us to determine if the electron will decay prior to exiting the radiative regime, i.e. $\tau_{e} < \tau_{c}$. Recalling the form of the time-dependent acceleration from Eqn. (4) we can find the deceleration time $\tau_{d}$, i.e. the maximum proper time that can elapse while still allowing the electron lifetime to be smaller than the critical time. Hence

\bqa
\tau_{c} &>& \tau_{\m} \frac{e^{2\pi / \tilde{a}}-1 }{1+ 5  \tilde{a}^{2} + 4 \tilde{a}^{4}} \non \\
 &>& \tau_{\m} \frac{\pi}{2} \frac{1}{ \tilde{a}^{5}} \non \\
 &>& \tau_{\m} \frac{\pi}{2} \frac{m_{\m}^{5}(\tau_{d}/\tau_{0} + 1 )^{10}}{ a_{0}^{5}}  \non \\ 
\Rightarrow \tau_{d} &<& \tau_{0}\lbk \frac{\tau_{c}}{\tau_{\m}} \frac{2}{\pi}\frac{E_{0}^{10}}{m_{e}^{10}}\frac{1}{m_{\m}^{5}x_{0}^{5}}  \rbk^{1/10} -\tau_{0}.
\eqa 

The above deceleration time evaluates to $\tau_{d} < 2.5\times 10^{-17} - \frac{6.7\times 10^{-16}}{E_{0}} $ s. For lifetimes less than the deceleration time, the electron will have the maximum probability to decay before it exits the radiative regime. Now, to analyze the lifetime of the electron we will look at the case of constant acceleration. As such we focus on the acceleration scale $a_{0}$. This will provide a first estimate for the electron lifetime and enable us to analyze the system using the initial electron energy $E_{0}$. We note $ m_{\m} m_{e}^{2}x_{0} = 55 \; \text{TeV}^{2}$. Then, we find $\tilde{a}_{0} = \frac{E_{0}^{2}}{x_{0}m_{e}^{2}m_{\m}} = \frac{E_{0}^{2}}{55}$. Note we are writing the initial energy in units of TeV. Our electron lifetime is then given by

\bqa
\tau_{e}(E_{0}) &=&   \tau_{\m} \frac{e^{2\pi 55/E_{0}^{2}}-1 }{1+ 5 \lb \frac{E_{0}^{2}}{55}  \rb^{2} + 4 \lb \frac{E_{0}^{2}}{55}  \rb^{4}}.
\eqa

In order to estimate if the electron decay has enough time to take place, we require the electron lifetime to be shorter than the deceleration time, i.e. the electron is statistically likely to decay before it leaves the radiative regime. Moreover, if the lifetime of the electron is smaller than the time scale $\tau_{0}$ than we can disregard the time-dependence of the acceleration in the analysis and use the acceleration scale $a_{0}$ to characterize the system. In Fig. 1 we plot all relevant time parameters which characterize the decelerating electron. It should be recalled that the lifetime of a decaying particle defines the characteristic time for an ensemble to lose $\frac{1}{e}$ of the initial population. The statistical nature of the decay should be kept in mind as we estimate the probability of decay based on the relevant parameters of the system.

\begin{figure}[H]
\centering  
\includegraphics[,scale=.9]{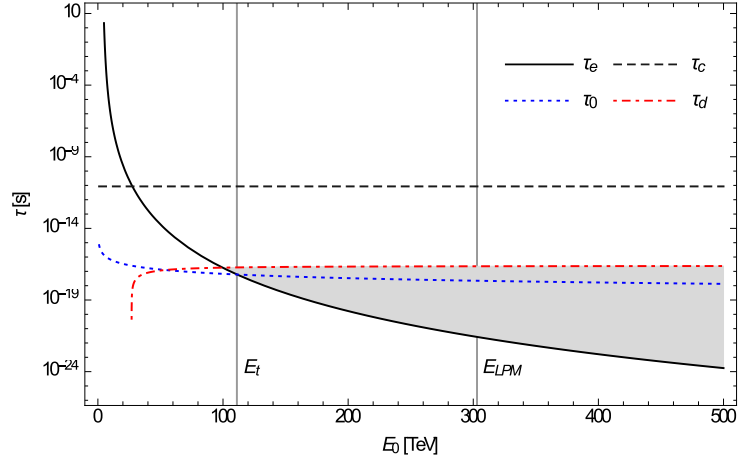}
\caption{Comparison of electron lifetime, critical time, deceleration time, and time scale as a function of the initial electron energy.}
\end{figure}

The shaded region of Fig. 1 indicates the regions where the lifetime is smaller than the deceleration time and we expect an electron decay to occur. Moreover, by inspection we see the energy scale at which the lifetime becomes smaller than the deceleration time is also very close to the energy scale time below which we may accurately approximate the acceleration as constant. We can determine this threshold energy $E_{t}$ for the lifetime to be shorter than the scale time $\tau_{0}$. Using a time-independent acceleration we find the threshold energy in the same way we computed the deceleration time in Eqn. (15). Thus, in the limit of high acceleration we have

\bqa
\tau_{0} &>& \tau_{\m} \frac{e^{2\pi / \tilde{a}}-1 }{1+ 5  \tilde{a}^{2} + 4 \tilde{a}^{4}} \non \\
&>& \tau_{\m} \frac{\pi}{2} \frac{1}{ \tilde{a}^{5}} \non \\
\frac{x_{0}m_{e}}{E_{0}}   &>& \tau_{\m}\frac{\pi}{2} \frac{ m_{\m}^{5} m_{e}^{10}x_{0}^{5}}{ E_{0}^{10}} \non \\ 
\Rightarrow E_{0} &> & \lbk \tau_{\m}\frac{\pi}{2} m_{\m}^{5} m_{e}^{9}x_{0}^{4} \rbk^{1/9}.
\eqa 

Then, for our electron in ice system, we find that for $E_{0} > 111.1$ TeV the acceleration is effectively constant while simultaneously satisfying the condition that the electron lifetime is less than the deceleration time. It is above this energy scale we expect to see an electron decay event. We can also calculate the final energy of the electron at the time of the decay using the proper velocity in Eqn. (4). Figure 2 contains the electron incident energies $E_{0}$, the energy at the time of decay $E_{f}$, and the change in energy $E_{d} = E_{0} - E_{f}$. At IceCube, the change in energy is, up to detector efficiency and effective volume, the measured energy that is deposited into the detector.

\begin{figure}[H]
\centering  
\includegraphics[,scale=1.1]{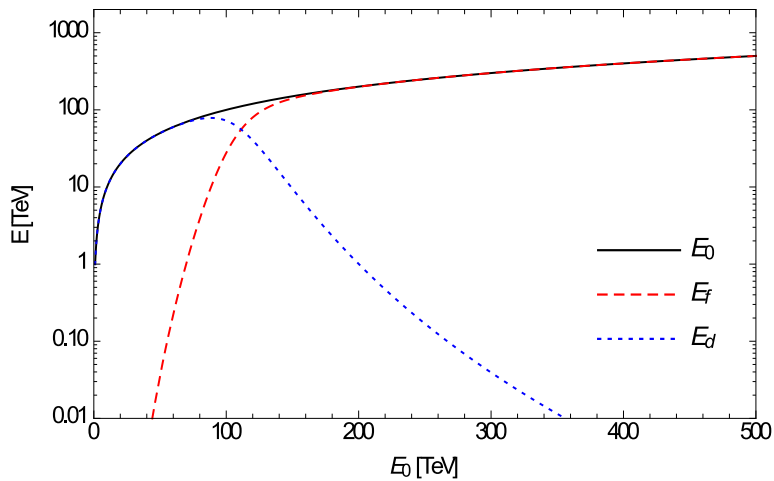}
\caption{Comparisons of the initial energy, final energy, and deposited energy in ice.}
\end{figure}

It is interesting to note that above the threshold energy $E_{t} \sim 111.1$ TeV the electrons do not lose any appreciable fraction of energy prior to converting to muons. However, at the $\sim 100$ TeV scale there will still be a measurable, and quite large, amount of energy deposited into the detector. We also note the condition that the time-dependent formalism used in the analysis remains valid is that $j/a^2 \ll 1$, where $j$ is the jerk [9]. We know the time-dependence of the acceleration becomes irrelevant for energies greater than $\sim 111.1$ TeV. However we must still verify the time-dependent formalism is valid to begin with. By taking the derivative of the proper acceleration we have $j = \frac{2j_{0}}{(\tau/\tau_{0}+1)^{3}}$ with $j_{0} = a_{0}/\tau_{0}$. Then the constraint implies

\bqa
\frac{j}{a^{2}} &\ll & 1 \non \\
\Rightarrow E_{0} &\gg & \frac{2m_{e}}{[1-\frac{2\tau}{\tau_{0}u_{0}}]}.
\eqa 

Pertaining to this analysis we recall that our lifetimes are much smaller than the scale time $\tau_{0}$. As such, we require $E_{0}  \gg 1.035$ MeV, which is most assuredly satisfied for all energies under consideration. Thus the condition necessary to ensure a valid time-dependent formalism has been verified. We close this section by commenting on any effect due to the sharp turn-on of the deceleration when the electron is produced [20]. For an acceleration turn-on time $\delta$, energy gap $\Delta E$, and lifetime $\tau$ the conditions for a thermal spectrum are $\tau \gg \delta $ and $ \tau \gg 1/\Delta E$. The first condition ensures that any transients due to the turn-on will be damped away and the second condition ensures that acceleration occurs long enough for the system to thermalize. In our analysis the turn-on time is given by the time it takes for the electron to traverse two scattering sites of ice. For ice, the lattice constant is $\ell_{s} \sim 2 \times 10^{-10}$ m and the turn-on time is given by $\delta = \ell_{s}/u_{0} = f_{s}\tau_{0}$. Note we have defined the fraction $f_{s} = \ell_{s}/x_{0} \sim 5 \times 10^{-10}$. As a function of the initial energy we have $f_{s}\tau_{0} \approx \frac{3\times10^{-25}}{E_{0}}$ s. Within the energy window analyzed here, $E_{0} \sim 100$ TeV, the turn-on time is given by $f_{s}\tau_{0} \approx 3\times10^{-23}$. From Fig. 1 we clearly see that $10^{-18} \; \text{s} \; \gtrsim \tau_{e} \gtrsim 10^{-22} \; \text{s}$ and therefore the condition that the lifetime is greater than the transient period, $\tau_{e}\gg f_{s}\tau_{0}$, is satisfied within the energy window where the decay is expected to occur. To ensure thermalization, we note the energy gap for the transition is given by the muon mass, $\Delta E = m_{\m}$. The condition, $\tau_{e} \gg 1/m_{\m}$, can easily be verified by noting $1/m_{\m} \sim 6 \times 10^{-24} \; \text{s}$. Therefore, based on the electron lifetime constraint from Fig. 1, the system has sufficient time to thermalize, i.e. $\tau_{e} \gg 1/m_{\m}$. Thus, for the acceleration-induced electron decay at IceCube, the conditions which are necessary to produce a thermal spectrum in the presence of a sharp turn-on are satisfied.      

\section{The Muon Spectrum}

Having determined that there is indeed an energy regime in which we expect an electron decay to occur, we now endeavor to compute the energy of the emitted muon. Although there are also neutrinos emitted in the decay process, we exclude them from the analysis on account of their leaving the detector with virtually zero interaction with the ice. The muons, however, are charged and will deposit energy, via bremsstrahlung, as they propagate away. Moreover, as we found in the first section, muons at IceCube energies will not have any appreciable deceleration in comparison to the electrons. The generalized spectra $\mathcal{N}$ from [9] is given by

\bqa
\mathcal{N}(\Delta E, a, n) &=& \frac{1}{\Gamma} \frac{G^{2}_{n}\tilde{\omega}}{(2 \pi)^{2}} \lb \frac{\Delta E+\tilde{\omega}}{\pi}\rb^{2n-3} \frac{1}{(4n-6)!!}  \prod_{k = 0}^{n-2}\lbk 1  + k^{2} \lb \frac{ a}{\Delta E+\tilde{\omega}} \rb^{2} \rbk  \frac{1}{e^{2\pi(\Delta E+\tilde{\omega})/|a|} - 1}.
\eqa

The parameter $\tilde{\omega}$ is the energy of the emitted particle, in this case the muon, as measured in a comoving frame instantaneously at rest with the accelerated particle, in this case the electron. Enforcing $n = 3$ for the electron excitation, and imposing the coupling $G^{2} = \frac{35}{8}\pi^{2}G^{2}_{f}$ we have

\bqa
\mathcal{N}_{\m}(a, \tilde{\omega}) &=& \tau_{e}(\tilde{a})\frac{35G^{2}_{f}\tilde{\omega} (m_{\m}+\tilde{\omega})^{3}}{1536 \pi^{3}}\frac{1+\lb \frac{a}{m_{\m}+\tilde{\omega}}  \rb^{2}}{e^{2\pi(m_{\m}+\tilde{\omega})/|a|} - 1}.
\eqa 

Then writing everything in terms of the initial energy $E_{0}$ we find

\bqa
\mathcal{N}_{\m}(E_{0},\tilde{\omega}) &=& \tau_{e}(E_{0})\frac{35G^{2}_{f}\tilde{\omega} (m_{\m}+\tilde{\omega})^{3}}{1536 \pi^{3}}\frac{1+\lb \frac{E_{0}^{2}}{m_{e}^{2}x_{0}(m_{\m}+\tilde{\omega})}  \rb^{2}}{e^{2\pi m_{e}^{2}x_{0}(m_{\m}+\tilde{\omega})/E_{0}^{2}} - 1}.
\eqa

Figure 3 contains spectra for various initial electron energies normalized to unity over $\tilde{\omega}$ via $N = \frac{\mathcal{N}}{\int \mathcal{N}d\tilde{\omega}}$. Recalling the threshold energy is $E_{t} = 111.1$ TeV, we will restrict our analysis to energies above the threshold so as to ensure the appropriate time-independent analysis. The Planckian spectra reflect the thermal nature of the accelerated reference frame.

\begin{figure}[H]
\centering  
\includegraphics[,scale=1.1]{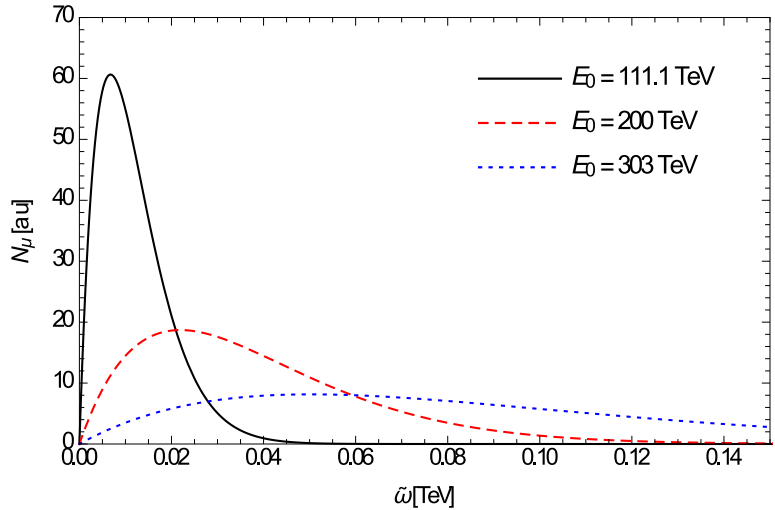}
\caption{The spectrum of the emitted muon for various initial electron energies.}
\end{figure}

The notoriously low statistics associated with neutrino interactions necessitates we examine the most probable energy emitted rather than the distribution. This is accomplished via determination of the peak energy of the emitted muon spectra using the generalized displacement law [9]

\bqa
\frac{xe^{x}}{e^{x} - 1} - \lbk \frac{1}{1-\frac{2 \pi \Delta E}{|a|x}} +(2n - 3)-2\lb \frac{2 \pi}{x}\rb^{2} \sum_{k = 0}^{n-2}\frac{k^{2}}{1  + k^{2} \lb \frac{2 \pi}{x} \rb^{2}}  \rbk &=& 0.
\eqa

The displacement coefficient $x$ is computed numerically and determines the peak muon energy via the expression

\bqe
\tilde{\omega} = x \frac{|a|}{2 \pi} - \Delta E.
\eqe

Here we find the inherently quantum mechanical energy of acceleration $E_{a} = \frac{xa\hbar}{2 \pi c}$. The forthcoming analysis shows there exists the potential to experimentally probe the nature of this energy. Now, application of the displacement law to our system necessitates enforcing the condition that $n=3$, $\Delta E = m_{\m}$, and $a = a_{0} = \frac{E_{0}^{2}}{m_{e}^{2}x_{0}}$. As such, we obtain

\bqa
\frac{xe^{x}}{e^{x} - 1} - \lbk \frac{1}{1-\frac{2 \pi 55}{E_{0}^{2}x}} +3-\frac{2}{\lb \frac{x}{2 \pi} \rb^{2}  +  1}  \rbk &=& 0.
\eqa

Numerically solving for $x$, for any incident electron energy, enables us to then determine the most probable muon energy as measured in the proper frame instantaneously at rest with the accelerated electron. It is worth mentioning that the muons energy comes from the accelerated temperature. The peak muon energy $\tilde{\omega}_{0}$ is given by

\bqa
\tilde{\omega}_{0} &=& m_{\m}\lbk  x \frac{E^{2}_{0}}{2 \pi 55} - 1\rbk.
\eqa

In Table. VI we have tabulated the displacement coefficient for various initial energies. We see that to within a percent we have $x \sim 1.8$. It is also interesting to note that the proper energy of the emitted muon grows as the square of the initial electron energy albeit weighted by a small prefactor. 

\begin{table}[H]
\centering
\begin{tabular}{ ccc}
\hline \hline \\ [-2.0 ex]
$E_{0}$~[TeV]~&~$x$~&~$\tilde{\omega}_{0}$~[GeV]~ \\
\hline \\ [-2.0 ex]
111.1 & 1.818 & 6.758 \\
\hline \\ [-2.0 ex]
200 & 1.800 & 21.92\\
\hline \\ [-2.0 ex]
303 & 1.796 & 50.33\\
\hline \\ [-2.0 ex]
\hline
\end{tabular}
\caption{The displacement parameter and associated peak muon energy for each incident electron energy.}
\end{table}

We can finalize this analysis by noting that emitted muon will also be boosted due to any residual velocity from the incident electron. We can now use the electron lifetime to compute the electrons proper velocity and thus the Lorentz gamma. We also compute the Lorentz gamma of the muon as measured in the electrons frame. This will enable us to determine the energy of the muon as measured in the lab frame. First, we know the muons energy, in the electrons frame, is $\tilde{\omega}_{0} = m_{\m}\tilde{\gamma}_{\m}$. From the muons Lorentz gamma we can also invert it to get the corresponding proper velocity. Hence

\bqa
\tilde{\gamma}_{\m} &=& \frac{\tilde{\omega}_{0}}{m_{\m}} \non \\
\Rightarrow \tilde{\beta}_{\m} &=& \lbk 1- \lb  \frac{m_{\m}}{\tilde{\omega}_{0}} \rb^{2}   \rbk^{1/2}.
\eqa

For the electron we know at the time of emission $\gamma_{e} = \frac{E_{f}}{m_{e}}$ and we can also solve for the electrons velocity $\beta_{e} = \sqrt{1-1/\gamma_{e}^{2}}$. Then using the composition of Lorentz gammas via the relativistic velocity addition formula [11] $\gamma_{\m} = [1 + \beta_{e}\tilde{\beta}_{\m}\cos{(\tilde{\theta}_{\m})}]\gamma_{e}\tilde{\gamma}_{\m}$, we have the muon energy, as measured in the lab frame, is given by $\omega = m_{\m}\gamma_{\m}$. The angle $\tilde{\theta}_{\m}$ is measured in the proper frame of the electron and determines the angle between the muons velocity relative to the electrons velocity. For both forward and backwards emitted muons we have 

\bqa
\omega_{\pm} &=& m_{\m}\gamma_{\pm} \non \\
&=& m_{\m}(1 \pm \beta_{e}\tilde{\beta}_{\m})\gamma_{e}\tilde{\gamma}_{\m} \non \\
&=&\tilde{\omega}_{0}\frac{E_{f}}{m_{e}} \lbk 1 \pm \lbk 1- \lb  \frac{m_{e}}{E_{f}} \rb^{2}   \rbk^{1/2}\lbk 1- \lb  \frac{m_{\m}}{\tilde{\omega}_{0}} \rb^{2}   \rbk^{1/2}\rbk.
\eqa  

Note, the plus or minus corresponds to muon velocity being parallel or anti-parallel to the electrons velocity respectively. These limits bound the muons energy when measured in the lab frame. Figure 5 contains a plot which outlines all emitted muon energies as a function of the electrons incident energy using the approximate displacement constant $x = 1.8$ for all energies. The shaded region to the right of the threshold energy $E_{t}$ partition comprises the domain of muon energies predicted by AQD. The forward and back scattered limits are denoted by $\omega_{+}$ and $\omega_{-}$ respectively. Regions above the initial electron energy are kinematically forbidden in Minkowski space and could provide the opportunity to investigate the nature of the energy of acceleration $E_{a}$. Below the initial electron energy both the standard model and AQD can describe the signal. Here an analysis comparing the energies of muons emitted by other processes would be needed to differentiate the signal. This region may also be used to compare the emission rates, and energies, of muons from the Unruh effect and from standard model radiative processes.

\begin{figure}[H]
\centering  
\includegraphics[,scale=.9]{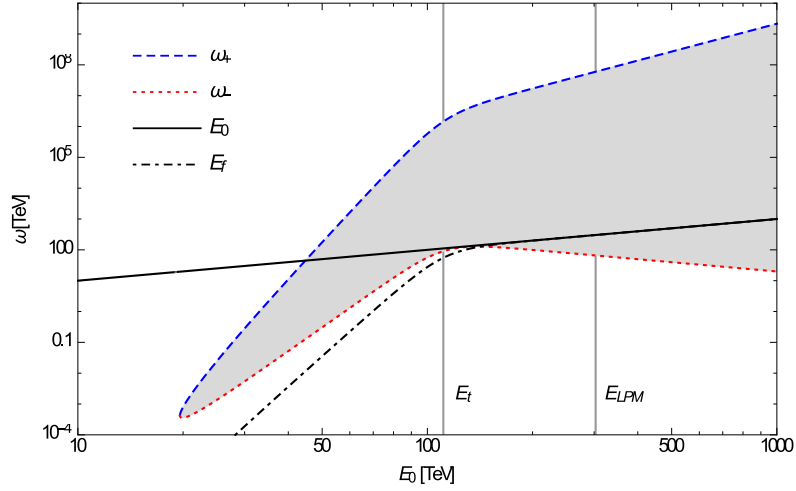}
\caption{Comparison of the initial electron energy and final muon energy as measured in the lab frame.}
\end{figure}

The parameter space of muon energies which kinematically conserve energy can be used to constrain the muon emission angle. For ultra relativistic electrons and muons $\beta_{e} \tilde{\beta}_{\m} \sim 1$, and the final lab frame muon energy $\omega_{\m} = E_{f}$ we have $ 1 = \frac{\tilde{\omega}_{0}}{m_{e}}[1+\cos{(\tilde{\theta}_{\m})}]$. Inverting the cosine then yields the muon emission angle. Hence

\bqa
\tilde{\theta}_{\m} &\approx & \pi - \lbk \frac{2m_{e}}{\tilde{\omega}_{0}}  \rbk^{1/2} \non \\
&= & \pi - \lbk \frac{2m_{e}}{m_{\m}\lb  x \frac{E^{2}_{0}}{2 \pi 55} - 1\rb} \rbk^{1/2}. 
\eqa

Thus for conservation of energy to be imposed, we require the emitted muon to be back scattered in the reference frame of the electron. This implies the muon is emitted away from the apparent horizon of the accelerated reference frame. After the muon has been emitted it will lose energy via a combination of ionizing and radiative processes [10]. This energy loss can be parametrized as

\bqa
\frac{d\omega}{dx} &=& -\alpha - \beta \omega \non \\
\Rightarrow \omega(x) &=& \omega_{0}e^{-x/x_{0}} + \frac{\alpha}{\beta}(e^{-x/x_{0}}-1).
\eqa

For muons in ice we have $x_{0} = 1/\beta = 3120$ m and $\alpha/\beta = .4762$ TeV. With these parameters, along with the monotonically increasing energy deposited by the muon $\omega_{d} = \omega_{0} - \omega(x)$, we can determine the initial energy of the muon to be

\bqa
\omega_{0} &=& \frac{\omega_{d}}{1-e^{-x/x_{0}}}-\frac{\alpha}{\beta}.
\eqa

The understanding of the energetics of both the electron and muon, in particular their deposited energies, enables one to confirm this acceleration-induced process. What is interesting to note is that there exists a rather appreciable parameter space where the muon has more energy than the initial electron. To conserve energy, a constraint is made on the muon emission angle which implies the muon is emitted away from the horizon. Moreover, any apparent violation of conservation of energy could be used to probe the energy of acceleration $E_{a} = \frac{xa\hbar}{2\pi c}$. The correlations between the emitted muon energy and the initial electron energy, along with a comparison with background muon and scattering emission rates, has the potential to distinguish the nature of the muon emission and thereby probe the existence of the Unruh effect. We must also mention the energy loss via bremsstrahlung will be suppressed at sufficiently high energies $E_{LPM}\gtrsim 303$ TeV due to the LPM effect [17]. Above this energy threshold, the electrons energy loss is no longer sufficient to produce the required acceleration and the electron decay probability goes to zero. 

\section{Electron decay signal at IceCube}
When a neutrino enters the IceCube detector it may interact with the ice via charged current or neutral current interactions. In neutral current interactions, a $Z$ boson is exchanged between the incident neutrino and a nucleon in the ice. The incident neutrino scatters off the nucleon and  and deposits energy in the form of a hadronic shower produced by the recoil of the nucleon. These signals have spherical topologies in the IceCube detector. In charged current interactions, a $W^{\pm}$ boson is exchanged between the incident neutrino and a nucleon in the ice. The neutrino produces the associated lepton in the final state along with a hadronic shower due to the nucleon recoil. The lepton produced then deposits its energy via a combination of radiative and ionization processes depending on the flavor. For electrons, the radiative energy loss is of such strength that the electron stops in the detector depositing an approximately spherical distribution of energy. For taus, the lifetime is so short that the decay and subsequent daughter products are all confined to the detector and deposits a spherical distribution of energy as well. The analysis here is done below the energy scale necessary to resolve the so called double bang signals of the creation and subsequent decay of the tau. These spherical signals are labeled as showers. For muons, the lifetime is of sufficient length that the muon leaves the detector depositing a track of energy deposited along the way. 

The signals are classified by their event topologies; either showers or tracks. For any incoming neutrino, one third of the event signals will be  due to neutral current interactions and the remaining two thirds of the event signals will be due to charged current interactions. For tau neutrinos, this implies all signals will have shower topologies. For muon neutrinos, one third of the events will have shower topologies while the remaining two thirds will have track topologies. For electron neutrinos one third of the events will have shower topologies. The remaining two thirds of the interactions which produce the associated electron will have shower topologies if the electron does not decay or track topologies if the electron decays into a muon. If we consider $N$ total incoming neutrinos species with the relative fractions $f_{e}$, $f_{\m}$, and $f_{\tau}$ such that $f_{e} + f_{\m} + f_{\tau} = 1$, we can classify the signal via the shower to track ratio $\frac{s}{t}$. In the absence of electron decay, the total showers $s$, tracks $t$, and the shower to track ratio $\sigma = \frac{s}{t}$ is given by

\bqa
s &=& f_{e}N + \frac{1}{3}f_{\m}N + f_{\tau}N \non \\
t &=& \frac{2}{3}f_{\m}N \non \\
\sigma &=& \frac{1}{2}\lbk 1 + \frac{3(f_{e} + f_{\tau})}{f_{\m}}    \rbk.
\eqa

To classify the event topologies in the presence of electron decay we denote each variable with a tilde, i.e. $\tilde{s}$, $\tilde{t}$, and $\tilde{\sigma} = \frac{\tilde{s}}{\tilde{t}}$. Then for electron decay we have

\bqa
\tilde{s} &=& \frac{1}{3}f_{e}N + \frac{1}{3}f_{\m}N + f_{\tau}N \non \\
\tilde{t} &=& \frac{2}{3}f_{e}N + \frac{2}{3}f_{\m}N \non \\
\tilde{\sigma} &=& \frac{1}{2}\lbk 1 + \frac{3 f_{\tau}}{f_{e} + f_{\m}}    \rbk.
\eqa

Finally, to probe the signal of electron decay, all one needs to analyze is the shower to track ratio at energies inside and outside of the electron decay window $E_{t} < E < E_{LPM}$. A difference in the measured ratios would warrant a more detailed search and, provided the flavour content is not energy-dependent, could be explained by electron decay. One could also use the measured ratios to pin down the incoming flavor content. Note Eqn.(31), Eqn. (32), and the unitary condition enable us to solve for the incoming flavors. Hence

\bqa
f_{e}&=& \frac{3\lb \sigma -  \tilde{\sigma} \rb}{2\lb 1+ \sigma  \rb  \lb 1 + \tilde{\sigma} \rb} \non \\
f_{\m}&=& \frac{3}{2\lb 1+ \sigma  \rb} \non \\
f_{\tau}&=& \frac{2\tilde{\sigma} - 1}{2 \lb 1 + \tilde{\sigma} \rb}.
\eqa

As an example we note that for incoming neutrino flavors of equal probability, i.e. $f_{e} = f_{\m} = f_{\tau} = \frac{1}{3}$, one would measure  $\sigma = \frac{7}{2}$ and $\tilde{\sigma} = \frac{5}{4}$. Another way to see this is simply as an excess of tracks with a sharp turn on at the threshold energy and a bit of a slower cutoff near the LPM energy. This analysis assumes there is no energy dependence in the relative fractions of the incoming neutrinos.

We close this section by commenting on the inclusion of fermions in the analysis rather than using scalar fields. By comparing the polynomials of multiplicity for an $n = 2$ fermionic description from [2] and an $n=3$ description used here, we note that the inclusion of fermions yield a higher order polynomial of multiplicity. The effect of this in the analysis would be to push the threshold energy $E_{t}$ to a lower energy scale. This would increase the electron decay window to encompass a larger number of neutrino signals at IceCube. We also must mention that the energy scales are consistent with a valid Fermi theory analysis. The fermionic computation developed in [5-7], adapted to electrons and muons, and without the use of an Unruh-DeWitt detector would provide a significantly more accurate prediction that satisfies all electroweak $t_{a_{0}}\ll m_{W^{\pm}}$, massless $a_{0} \gg m_{i}$, time-independent $t_{e} \ll \tau_{0}$, and perturbative approximations $G^{2} < 1$.

\section{Conclusions}
In this paper we utilized the radiative deceleration of electrons in ice to induce their decay back into muons. The acceleration profile, in the radiative regime, was developed and the relevant energy scales were tabulated for electrons and muons in different media including ice. We computed the electron lifetime and showed that a time-independent acceleration was sufficient for analysis beyond an accessible threshold energy. We also showed that above the energy threshold the electron lifetime was smaller than the critical time to exit the radiative regime and thus had an appreciable probability of decaying. The muon spectrum was computed in the electrons proper frame and also boosted to the lab frame. Using the radiative energy loss of both the electron and muon, we also computed the energy deposited in the ice by both particles. The analysis was carried out at IceCube energies. A description of the signal topologies in the presence of electron decay was also included. The resulting analysis shows that IceCube has the potential to verify the formalism of AQD, provide experimental evidence for the Unruh effect, investigate the nature of the quantum mechanical energy of acceleration $E_{a}$, and pin down the incoming neutrino flavor content.

\section*{Acknowledgments}
The author is indebted to George Matsas for proofreading this manuscript and wishes to thank Luiz da Silva and Niayesh Afshordi for many valuable discussions. This research was supported, in part, by the Leonard E. Parker Center for Gravitation, Cosmology, and Astrophysics, the University of Wisconsin-Milwaukee Department of Physics, and the Perimeter Institute for Theoretical Physics.

\goodbreak

\end{document}